\documentclass[astrosymb,trackchanges,twocolumn]{aastex701}
\usepackage{xurl}
\hypersetup{linkcolor=red,citecolor=blue,filecolor=cyan,urlcolor=magenta}

\usepackage{amsmath}
\usepackage{subfigure}
\usepackage{amsmath}
\usepackage{multirow}
\usepackage{booktabs}
\usepackage{graphicx,natbib}
\usepackage{bm}
\usepackage[normalem]{ulem}
\usepackage{xcolor}
\usepackage[normalem]{ulem}

\shorttitle{Speckle Interferometry with the DOT}
\shortauthors{Km N. Rai et al.}

\received{\today}
\revised{\today}
\accepted{November 30, 2026}

\begin{document}

\title{Stellar Multiplicity via Speckle Interferometry and Bayesian Inference with the 3.6 m Devasthal Optical Telescope}

\author[orcid=0000-0002-9939-805X, sname='Rai']{Km Nitu Rai}
\affiliation{Aryabhatta Research Institute of Observational Sciences (ARIES), Manora Peak, Nainital 263001, India}
\email{niturai201296@gmail.com}  

\author[orcid=0000-0002-0151-2361, sname='Panwar']{Neelam Panwar} 
\affiliation{Aryabhatta Research Institute of Observational Sciences (ARIES), Manora Peak, Nainital 263001, India}
\email{neelam@aries.res.in}

\author[orcid=0000-0002-4331-1867, sname='Pandey']{Jeewan C. Pandey} 
\affiliation{Aryabhatta Research Institute of Observational Sciences (ARIES), Manora Peak, Nainital 263001, India}
\email{jeewan@aries.res.in}

\author[orcid=0000-0003-2042-5762, sname='Kumar']{T. S. Kumar} 
\affiliation{Aryabhatta Research Institute of Observational Sciences (ARIES), Manora Peak, Nainital 263001, India}
\email{kumar@aries.res.in}

\author[orcid=0000-0001-6500-9987, sname='Sarangi']{Subrata Sarangi} 
\affiliation{Visiting Associate, Inter-University Centre for Astronomy and Astrophysics, Ganeshkhind, Pune 411007, Maharashtra, India.}
\email{ssarangi.itech@gmail.com}

\author[orcid=0000-0003-0136-2153, sname='Saha']{Prasenjit Saha} 
\affiliation{Physik-Institut, University of Zurich, Winterthurerstrasse 190, 8057 Zurich, Switzerland}
\email{psaha@physik.uzh.ch}

\begin{abstract}
Conventional ground-based telescopes are fundamentally limited by atmospheric turbulence, which degrades their angular resolution. Speckle interferometry (SI), based on short-exposure imaging, mitigates these effects and recovers diffraction-limited information. In this work, we implement SI using autocorrelation techniques to resolve binary stars and introduce a Bayesian framework to model the speckle observations and estimate binary parameters. Speckle observations of 52 Ori in the white light and 10 Ari in the SDSS $g$-band were obtained using the 3.6-m Devasthal Optical Telescope (DOT) at Aryabhatta Research Institute of Observational Sciences (ARIES), Nainital. The conventional Fourier analysis provides angular separations and position angles that are consistent with previously published measurements. The Bayesian analysis additionally provides posterior estimates of the apparent magnitude difference, angular separation, and position angle. For 52 Ori, the inferred parameters are $0.39 \pm 0.02$ mag, $1.062'' \pm 0.007''$, and $220.3^\circ \pm 0.4^\circ$, whereas for 10 Ari, the corresponding measured values are $2.55 \pm 0.07$ mag, $1.527'' \pm 0.005''$, and $345.9^{\circ} \pm 0.2^\circ$. These results demonstrate the feasibility of SI with the 3.6-m DOT and the potential of Bayesian modeling as a complementary approach to binary-star characterization. Further improvements in detector characterization, calibration, and noise modeling are required to assess the accuracy and robustness of the Bayesian parameter estimates.
\end{abstract}


\section{Introduction}
The Earth's atmosphere poses a significant challenge to high-spatial-resolution astronomical imaging. Due to vast interstellar distances and the effects of Earth's atmospheric turbulence, stellar objects appear as unresolved point sources to the naked eye and as blurred images through ground-based optical telescopes. Even large-aperture telescopes are fundamentally limited in achieving their theoretical diffraction-limited resolution.

To overcome these limitations, several high-resolution imaging techniques have been developed over the past few decades. These include adaptive optics \citep{hardy1998adaptive, tyson2022principles}, lucky imaging \citep{fried1978probability}, and various interferometric methods such as speckle masking \citep{lohmann1983speckle} and Speckle Interferometry \citep[SI;][]{Dainty1975, scott2018nn}. Among these, SI has proven particularly effective for achieving near-diffraction-limited resolution with ground-based telescopes.

First introduced by \cite{labeyrie1970attainment}, SI is theoretically grounded in the phenomenon of photon correlations in pseudo-thermal radiation from stellar objects and established by earlier works such as \cite{1957RSPSA.242..300B, Glauber1963PhotonCorrelations}. It captures a series of short-exposure images of the target system, on the order of milliseconds, thus effectively ``freezing'' the atmospheric turbulence. These exposures exhibit high-frequency spatial information as speckle patterns. However, these patterns also capture low-frequency structural details of the star and the stellar system, as in the case at hand. These low-frequency details, carrying important source information, can be recovered through post-processing of these short-exposure images using standard Fourier Transform (FT) methods \citep{labeyrie1970attainment, butters1971double, goodman2007speckle}. The number of objects in the stellar system can also be inferred from the autocorrelation of these images \citep{goldfischer1965autocorrelation, goodman2007speckle}. However, this method does not apply to full-image reconstruction, as the phase information is lost in the calculation of the power spectrum (PS). To overcome this limitation, higher-order statistical techniques, such as speckle masking, triple correlation, and bispectrum analysis, are required, as such techniques retain phase information and enable diffraction-limited image reconstruction \citep{1977OptCo..21...55W, weigelt1983image, lohmann1983speckle}. The speckle masking method was initially developed to reconstruct the exact image of the objects \citep{1977OptCo..21...55W}. It applied a masking vector to each speckle frame before processing, which serves as a synthetic reference star. Later, a triple correlation was applied to it along with the original speckle frame \citep{weigelt1983image}. The FT of this correlation, also called the bispectrum, returns the exact image of stellar objects and their relative position according to the chosen masking vector \citep{lohmann1983speckle}.

Originally developed to resolve bright binary systems \citep{1989AJ.....97..510M}, SI has, since its initial demonstrations \citep{labeyrie1970attainment}, been successfully applied worldwide across a wide range of astrophysical applications. These include the detection and orbital characterization of binary and higher-order multiple stellar systems, the study of Wolf–Rayet stars in close binaries, and the identification of previously unknown companions in stellar systems \citep{mcalister1977speckle, weigelt1986eta, hartkopf1999iccd, horch2017, howell2024high, Howell_2025}. State-of-the-art instruments at telescopes such as the 3.5-meter WIYN Observatory \citep{scott2018nn}, the 4.1-meter SOAR Observatory \citep{2008PASP..120..170T, tokovinin2024speckle}, and the 8.1-meter Gemini telescopes \citep{scott2021twin} have successfully used SI to characterize intermediate binaries with a relative separation of less than $1.1''$. Most recently, SI observations with Gemini North led to the detection of a second companion around Betelgeuse \citep{Howell_2025}, highlighting the capability of SI as a high-resolution technique for resolving multiple stellar systems. \cite{saha1999development} developed a speckle interferometer for the 2.38-m Vainu Bappu telescope and measured the angular separation of HR 4891 and HR 5298 as $0.094''$ and $0.109''$, respectively \citep{saha2002speckle}.

In this work, we have used the 3.6 m Devasthal Optical Telescope \citep[DOT;][]{2018BSRSL..87...29K} to obtain speckle observations of binary systems and to resolve them using the SI technique. DOT has a theoretical resolving power of approximately $0.035''$ at $\sim$ 500 nm. The typical sub-arcsecond median seeing limits the telescope’s ability to resolve binary systems with small angular separations. Imaging observations of binary stars conducted with a test camera on the DOT in December 2015 demonstrated a minimum resolvable angular separation of approximately $0.37''$ under optimal seeing conditions \citep{sagar20193, Sagar2020}. In the present study, we have observed two binary systems, 52 Ori (52 Orionis) and 10 Ari (10 Arietis), which exhibit angular separations of roughly $1.0''$ and $1.5''$, respectively. An angular separation of more than $1''$ can be resolved with typical medium-sized classical optical telescopes. However, this study lays the groundwork for an SI-based pipeline for DOT, with the goal of eventually expanding it to all major telescopes at ARIES.

This paper is organized as follows. Section 2 sketches the basic theoretical concepts underlying the SI technique. Section 3 presents an outline of the Bayesian Model of the amplitude spectrum (AS) of the short-exposure images and introduces other estimated and relevant system priors. Section 4 provides basic information on the 3.6-m DOT and details of the speckle observations. Section 5 describes the two target stellar systems and presents, in sequence, the results of this pilot study. In Section 6, we summarize the results obtained for both binary systems and discuss their implications. We also highlight the prospects of SI as a backend instrument for the DOT and, more broadly, for other optical telescopes in India.
\section{Theoretical Background for Speckles}\label{sec:theory}
Ground-based optical observations are fundamentally limited by time-dependent and high-frequency random phase distortions in the incoming wavefronts from stellar sources. However, over very short timescales, typically a few tens of milliseconds, the phases across the arriving wavefront can be considered constant. Signals recorded with a high-speed detector within such intervals capture barely discernible coherence patterns and diffraction-limited information encoded in the speckles. Subsequent analysis of speckle frames recorded over a short time interval enables the recovery of diffraction-limited details and, in the case of binaries, the determination of binary parameters.
\begin{figure*}[hbt]
	\centering
	\includegraphics[width=0.49\linewidth]{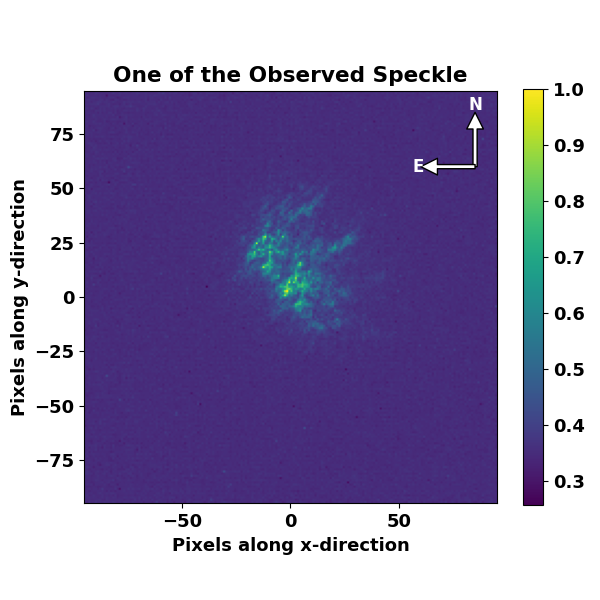}\hfil
	\includegraphics[width=0.49\linewidth]{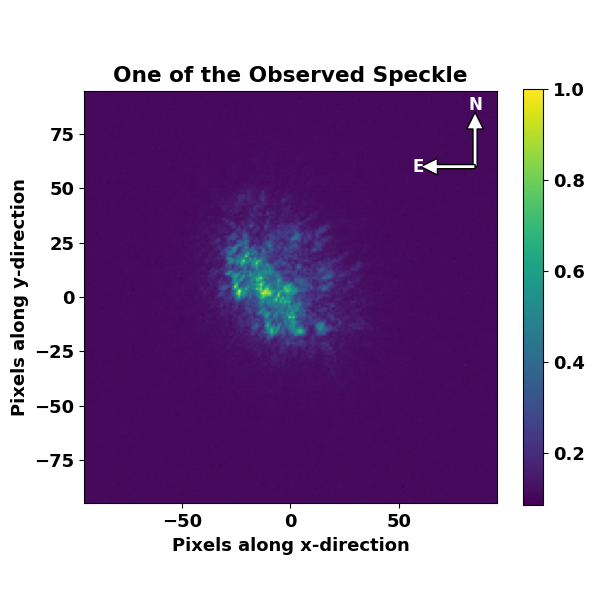}
	\caption{The left and right panels show representative speckle patterns of 52 Ori recorded with exposure times of 2 ms and 10 ms, respectively. The displayed images correspond to the region of interest (ROI) extracted from the raw images. The color bar indicates the normalized intensity distribution across the pixels.}
	\label{fig:Ori}
\end{figure*}
\begin{figure*}[hbt]
	\centering
	\includegraphics[width=0.49\linewidth]{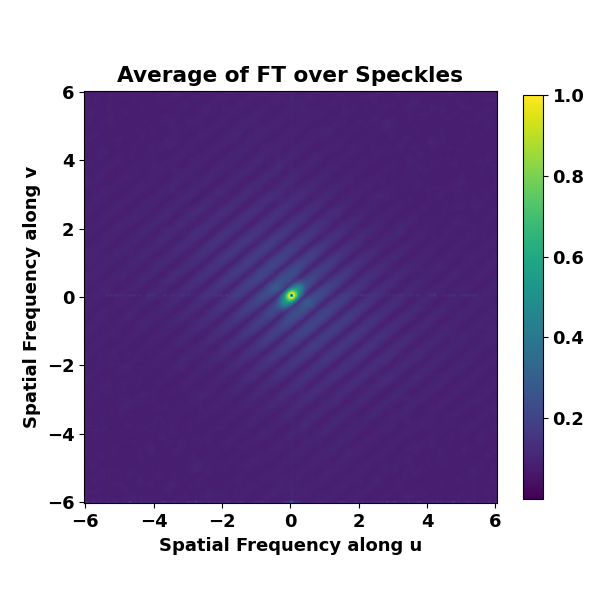}\hfil
	\includegraphics[width=0.49\linewidth]{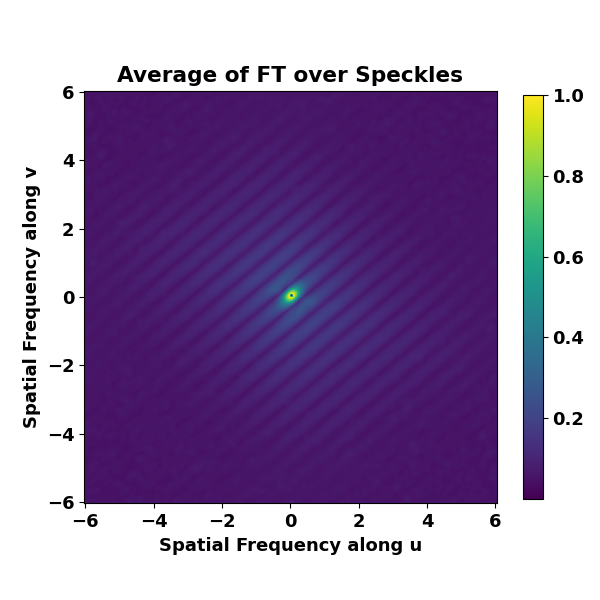}
	\caption{Fringe patterns of 52~Ori obtained by averaging the Fourier transforms (AS) of short-exposure image frames, such as shown in Fig.~\ref{fig:Ori}.  The left panel here is obtained from 1500 frames with exposure times of 2~ms per frame, while the right panel is obtained from 1000 frames of 10~ms each. The $\rm pixel^{1/2}$ of both datasets is shown here for clarity. The colorbar represents the normalized fringe distribution of 52 Ori over pixels.}
	\label{fig:Ori_FT}
\end{figure*}
\begin{figure*}[hbt]
	\centering
	\includegraphics[width=0.49\linewidth]{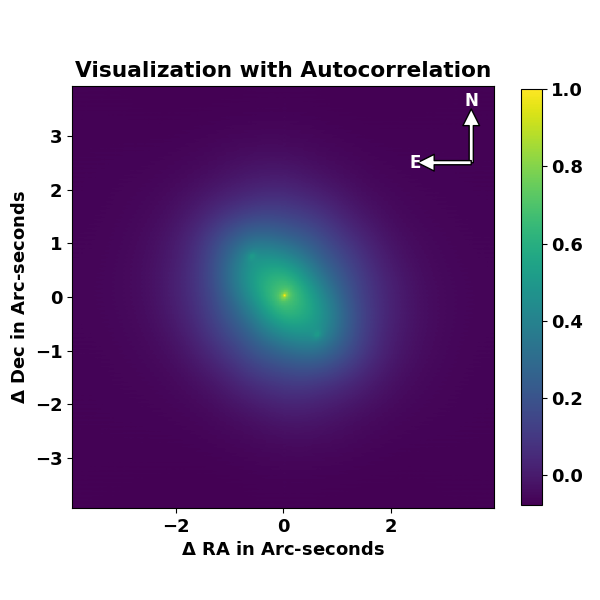}\hfil
	\includegraphics[width=0.49\linewidth]{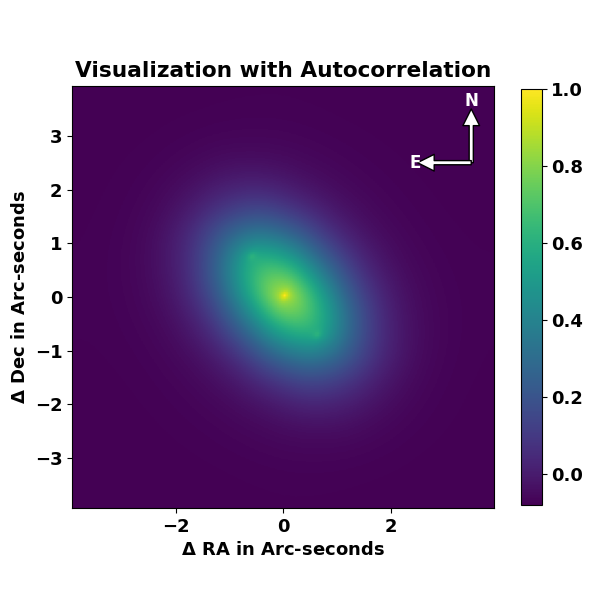}
	\caption{ Auto-correlation of frames in the system 52~Ori --- the inverse Fourier transform of Fig.~\ref{fig:Ori_FT}, so to speak.  The colorbar represents the autocorrelation amplitude over pixels, normalized to the peak value at the origin. The effect of the atmosphere on the PSF is evident as the exposure time of individual frames increases from 2~ms to 10~ms: the left panel shows clearer spots than the right panel.}
	\label{fig:Ori_auto}
\end{figure*}
\begin{figure*}[hbt]
    \centering
    \includegraphics[width=0.49\linewidth]{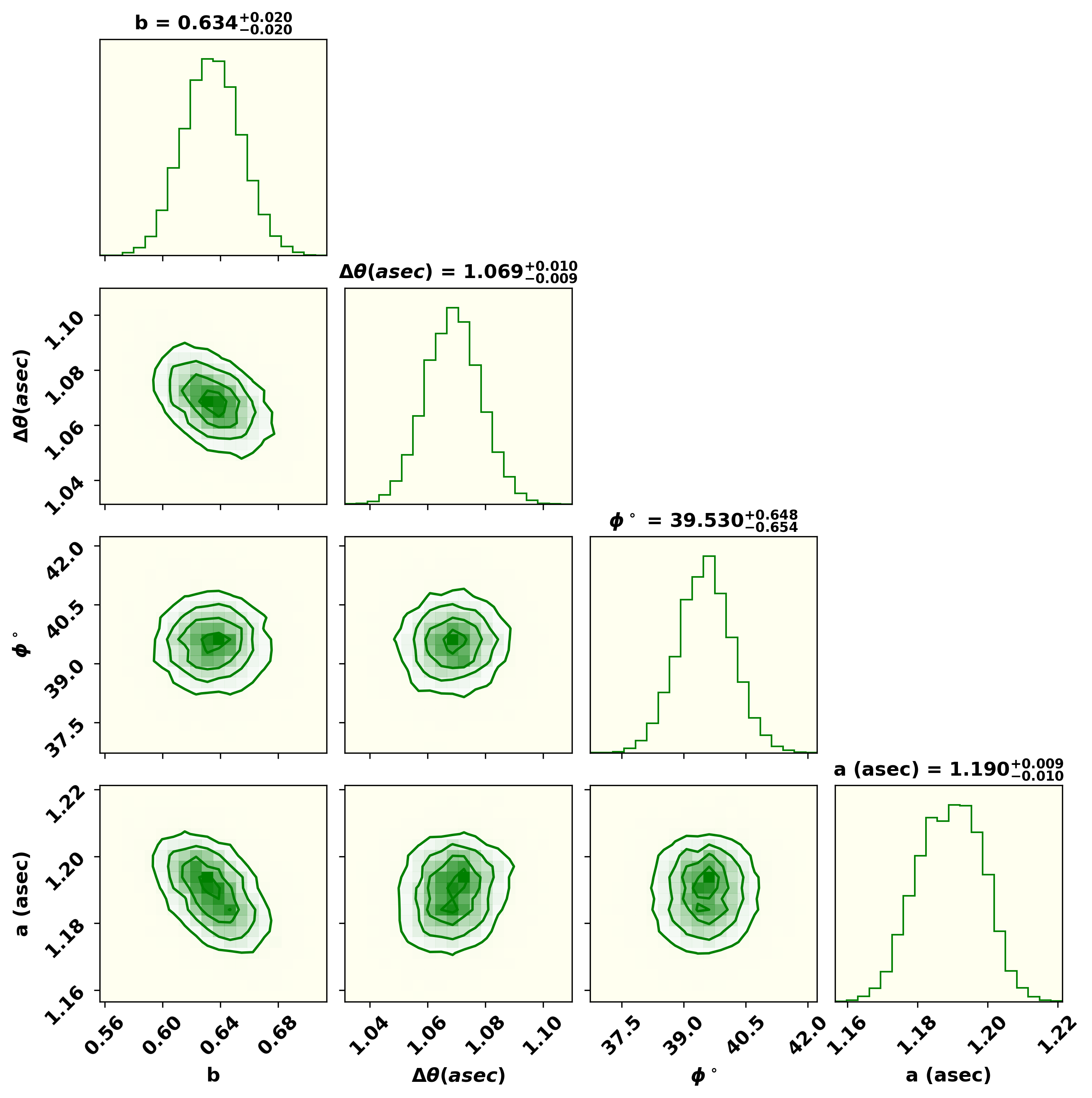}\hfil
	\includegraphics[width=0.49\linewidth]{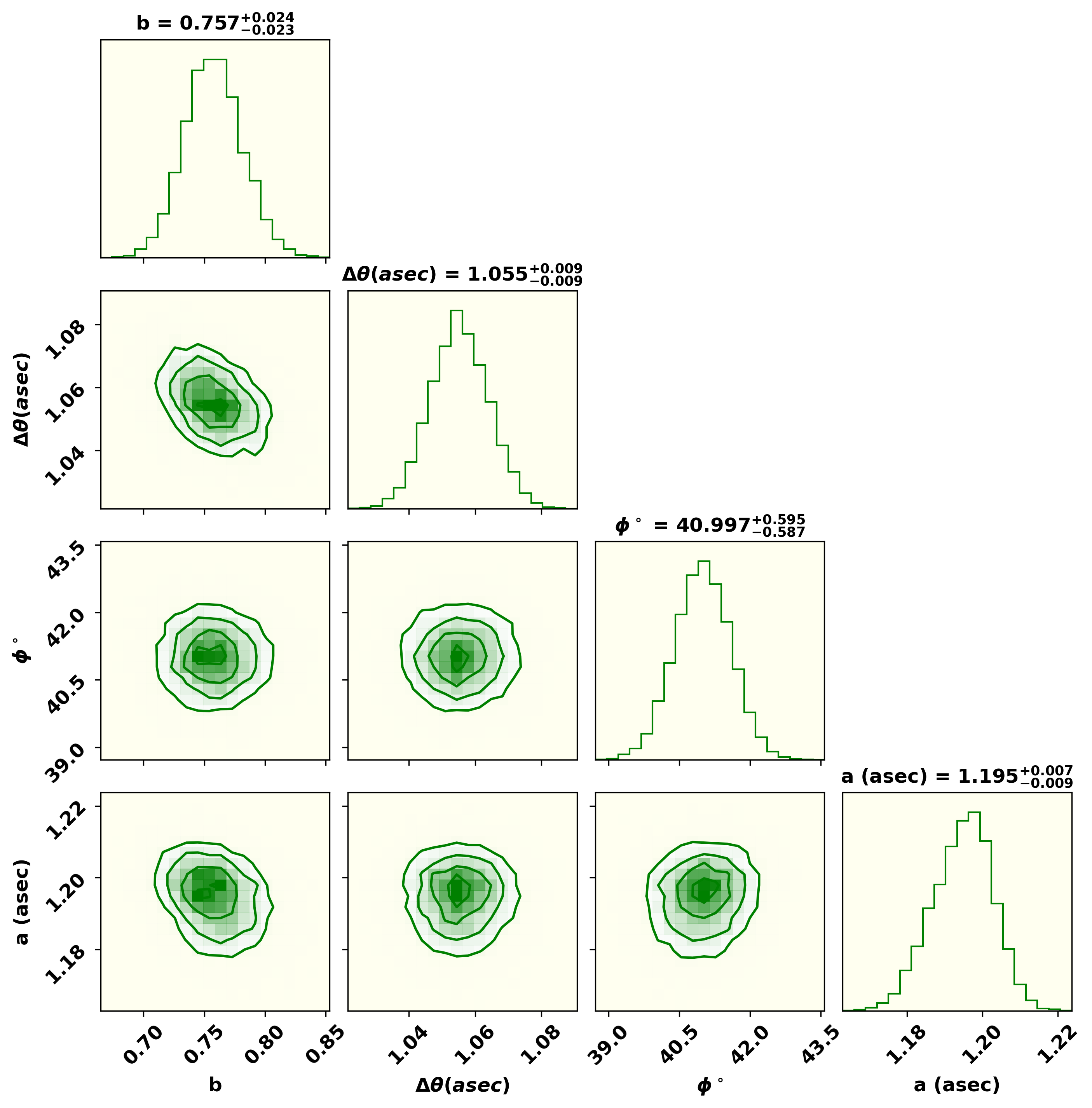}
	\caption{The left and right panels show the results of parameter estimation of 52 Orionis using the Bayesian modeling of the AS (Fig.~\ref{fig:Ori_auto}) of speckle patterns for 2 ms and 10 ms exposure times  (one example shown in Fig.~\ref{fig:Ori}), respectively.}
	\label{fig:Ori_para}
\end{figure*}

\begin{figure*}[hbt]
	\centering
	\includegraphics[width=0.49\linewidth]{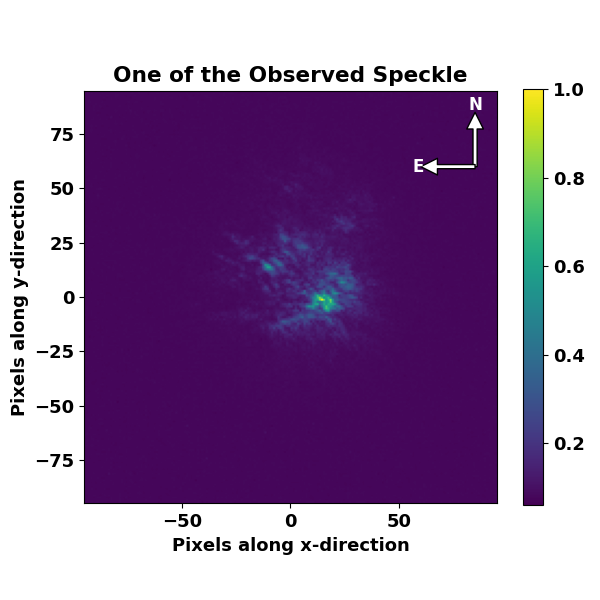}\hfil
	\includegraphics[width=0.49\linewidth]{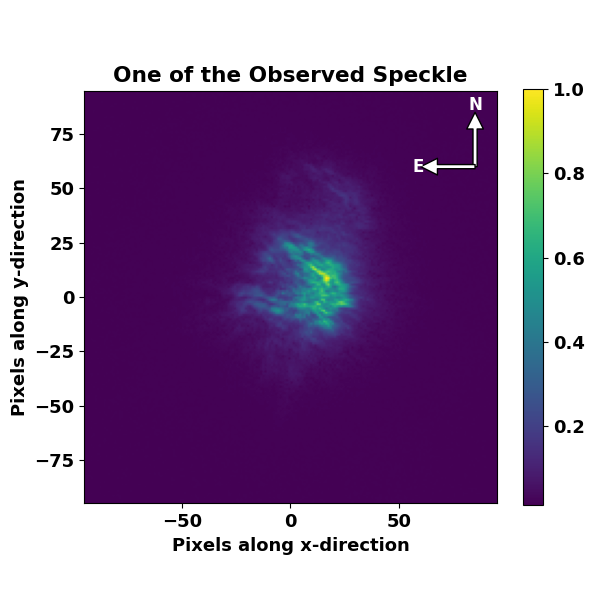}
	\caption{The left and right panels show the speckle patterns of 10 Ari with 10 ms and 100 ms exposure times, respectively. The colorbar represents the normalized intensity distribution over pixels.}
	\label{fig:Ari}
\end{figure*}
\begin{figure*}[hbt]
	\centering
	\includegraphics[width=0.49\linewidth]{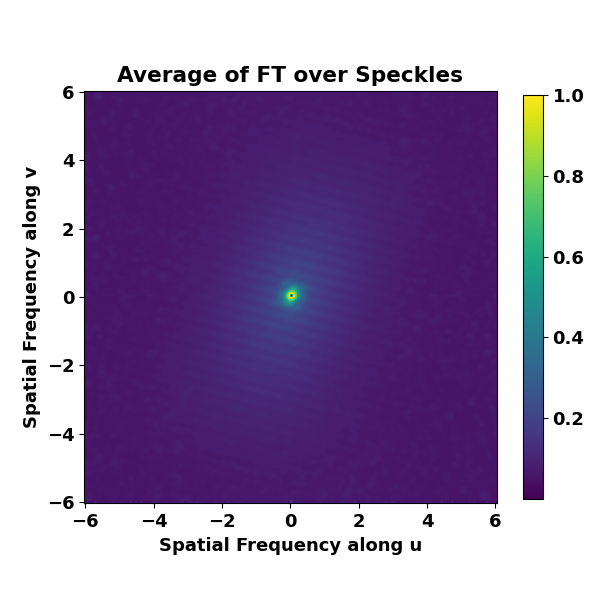}\hfil
	\includegraphics[width=0.49\linewidth]{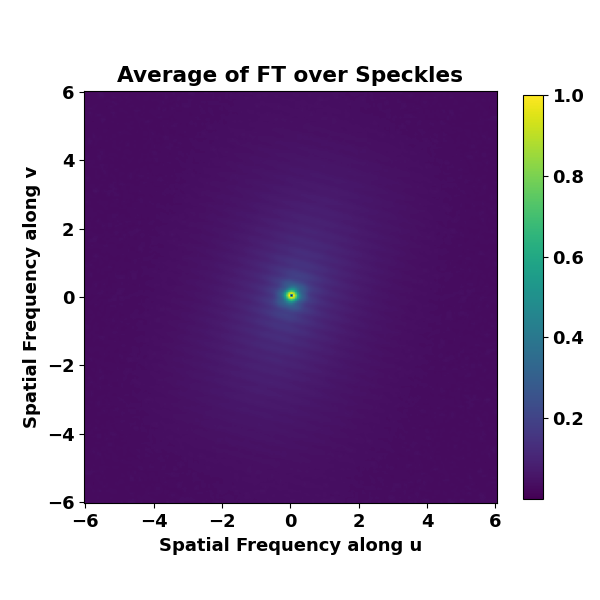}
	\caption{The left and right panels show the fringe pattern of 10 Ari obtained by averaging over the Fourier Transforms (AS) of 1000 short-exposure image frames with exposure times of 10 ms and 100 ms per frame, respectively, as shown in Fig.~\ref{fig:Ari}. As the companion of 10 Ari is very faint, we have taken the pixel$^{(1/2)}$ here for clear visualization. The colorbar represents the normalized fringe distribution of 10 Ari over pixels.}
	\label{fig:Ari_FT}
\end{figure*}
\begin{figure*}[hbt]
	\centering
	\includegraphics[width=0.49\linewidth]{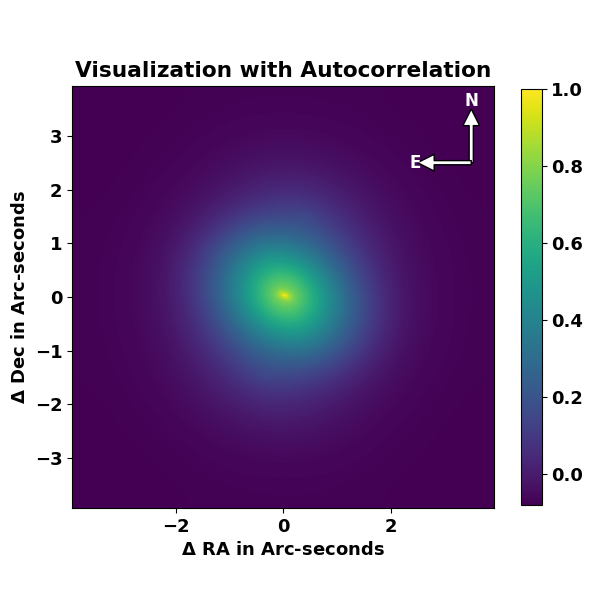}\hfil
	\includegraphics[width=0.49\linewidth]{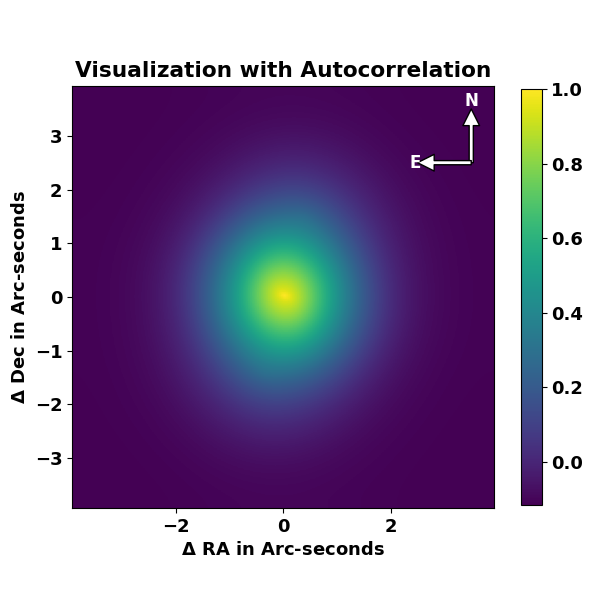}
	\caption{The left and right panels show the graphical representation of 10 Ari after taking the inverse Fourier transform of the average PS, respectively. However, the side dots are not visible clearly while having the fringe nature in the diffraction pattern shown in Fig.~\ref{fig:Ari_FT}. Here also, the effect of exposure time on the PSF is evident: the left panel appears less blurred than the right one. The colorbar represents the autocorrelation amplitude over pixels, normalized to the peak value at the origin.}
	\label{fig:Ari_auto}
\end{figure*}

\begin{figure*}[hbt]
	\centering
	\includegraphics[width=0.49\linewidth]{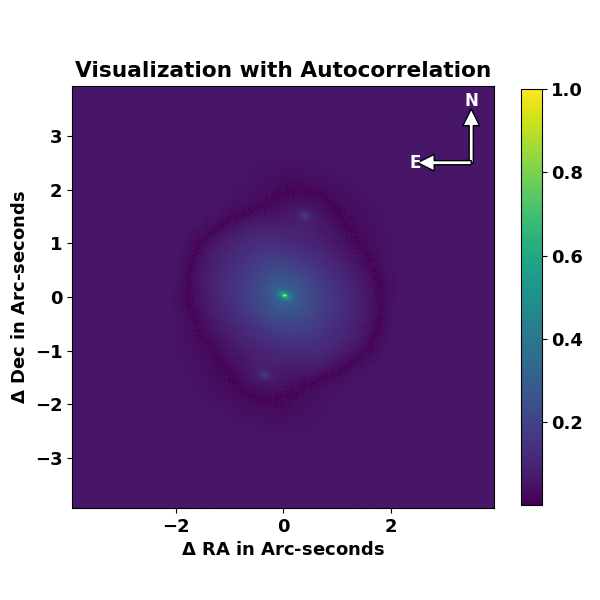}\hfil
	\includegraphics[width=0.49\linewidth]{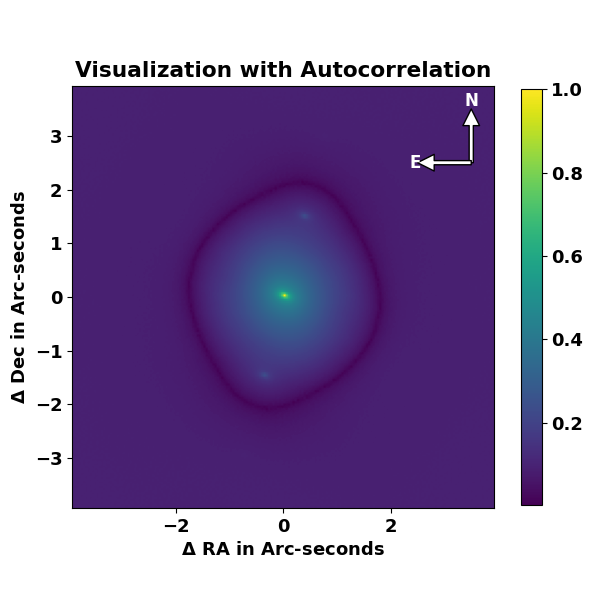}
	\caption{The left and right panels show the graphical representation of 10 Ari after taking the IFFT of the AS, respectively. The $\rm pixel^{1/2}$ has been taken over the image for clear visualization of secondary dots.}
	\label{fig:Ari_auto_PS}
\end{figure*}

\begin{figure*}[hbt]
    \centering
    \includegraphics[width=0.49\linewidth]{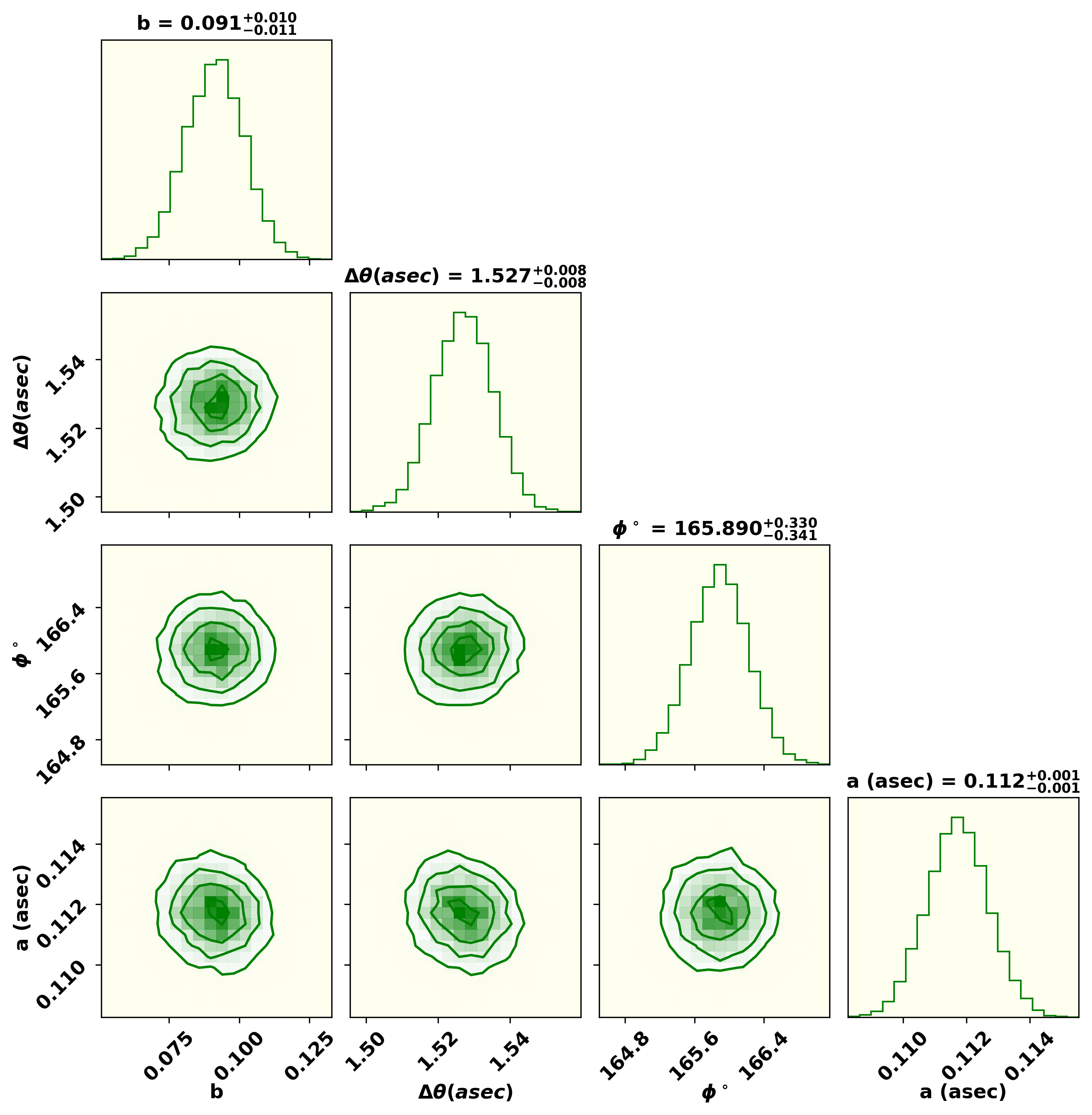}\hfil
    \includegraphics[width=0.49\linewidth]{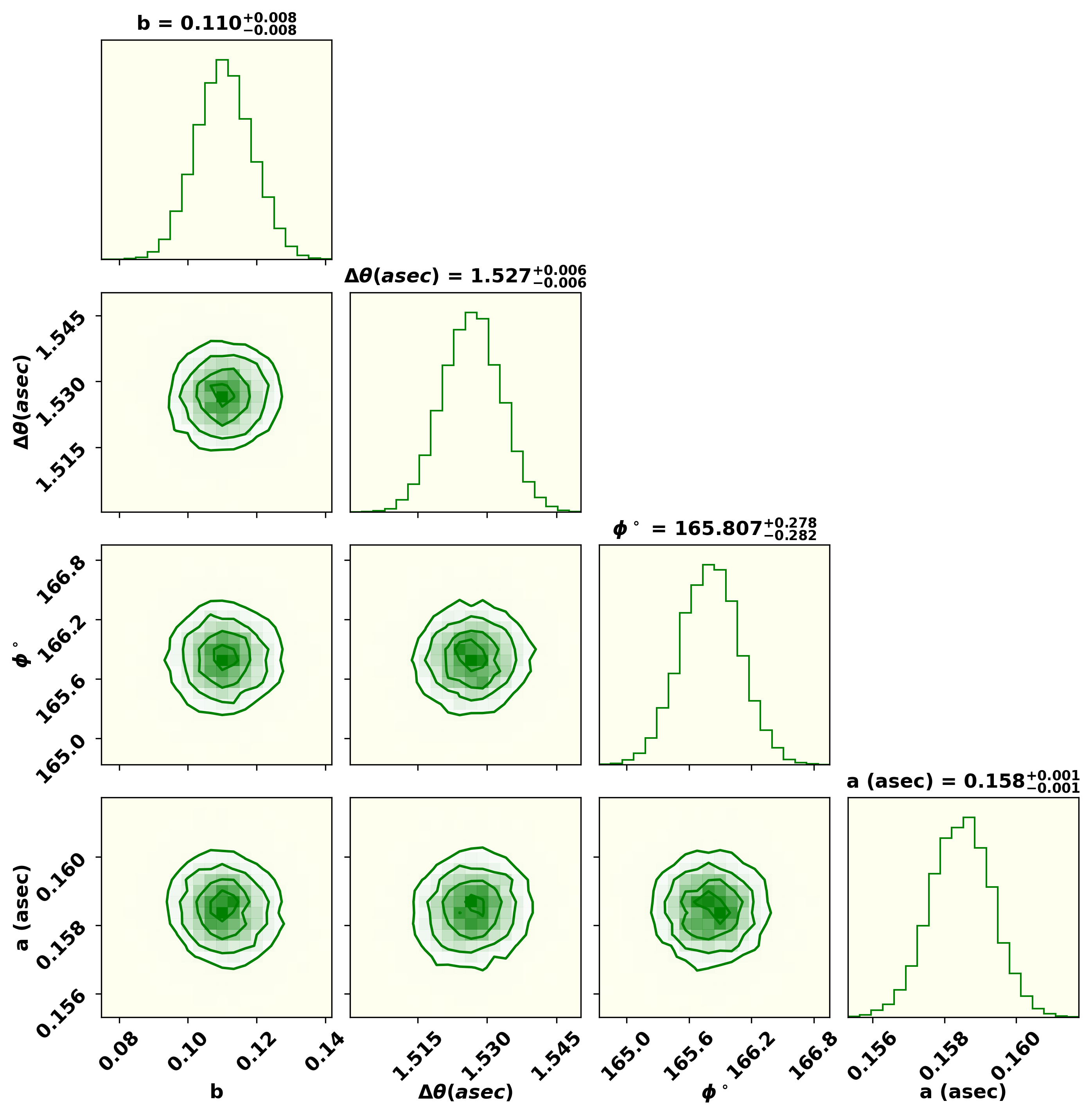}
    \caption{The left and right panels show the results of parameter estimation for 10 Ari using Bayesian modeling of the IFFT of the AS (shown in Fig.~\ref{fig:Ari_auto_PS}) of speckle patterns for 10 ms and 100 ms exposure times per frame, respectively.}
	\label{fig:Ari_para}
\end{figure*}
\subsection{Diffraction Limit and Atmospheric Constraints}
The diffraction-limited angular resolution at wavelength $\lambda$ for a telescope of objective diameter $D$ is given by
\begin{equation}
 \theta \approx 1.22 \frac{\lambda}{D}.
\end{equation}
In the absence of the Earth's atmospheric turbulence, observation of a single star yields a circular Airy disk, whereas a binary star system should produce interference fringes modulated on the Airy pattern. In practice, atmospheric turbulence distorts the incoming wavefront and blurs the Airy pattern, thereby preventing direct observation of fine structure.

\subsection{Fourier Analysis of Speckle Patterns}\label{sec:FT}
SI attempts to circumvent atmospheric blurring by recording short-exposure frames of the wavefronts reaching the objective of the telescope. Each frame contains a random speckle pattern along with the source information. The FT of the frame retains the low-frequency information of the stellar object while treating the atmospheric perturbations as multiplicative high-frequency noise.
Since each frame contains only a partial sampling of the photon distribution, the Fourier-domain representation of a single speckle image yields blurred array patterns that also vary frame-to-frame. However, by averaging the FT over many short-exposure frames, the overall effect of atmospheric noise diminishes, and the object’s low spatial-frequency content becomes more clearly discernible. For binary stars, this averaging recovers characteristic fringe patterns in the Fourier domain.

\subsection{Autocorrelation: Graphical Representation}
Following the Wiener–Khintchine theorem \citep{khintchine1934korrelationstheorie}, the ensemble-averaged PS (squared modulus of the FT) of the speckle frames can be inverse Fourier-transformed to yield the autocorrelation of the object’s diffraction-limited image. For a binary star, the autocorrelation function displays a central bright spot accompanied by two symmetrically located secondary peaks corresponding to the projected separation of the components. The presence of these symmetric secondary peaks reflects the intrinsic 180-degree ambiguity of the autocorrelation. The projected separation is determined by measuring the distance between the central peak and either of the secondary peaks.
This provides a direct graphical signature of the target's binary nature. 
While this method extracts projected separations and position angles, fully reconstructing the image requires higher-order techniques such as triple correlation or bispectrum analysis.
\section{Modeling of Amplitude Spectrum}\label{sec:model}
Conventional diffraction-limited imaging techniques provide point estimates of parameters. However, a full probabilistic distribution of the parameters can be obtained by applying a statistical method to the speckle data. It also helps constrain additional information on the brightness ratio and atmospheric parameters. We employ a Bayesian inference model on the average of the Fourier transform of the speckles, called the amplitude spectrum (AS). It is calculated over observed frames using the FT technique. 

From the standard theory of Fraunhofer diffraction, the intensity distribution $I(u)$ (maximum at $u=0$) on Earth for a single star is given by
\begin{equation}
    I(u) = I_{1} \left[2\frac{J_1(u)}{u}\right]^2
\end{equation}
where $J_1$ represents the first-order Bessel function of the first kind, $I_{1}$ is the brightness of star and $u$ is defined as
\begin{equation}
    u = \frac{\pi D \theta}{\lambda} = \frac{\pi \theta}{a}
\end{equation}
with observational wavelength $\lambda$ and telescope diameter $D$. In this model, the scale parameter $a=\lambda/D$ depends on the observational wavelength employed by the telescope. Besides, it also incorporates the diffraction limit, along with atmospheric and instrumental effects, as characterized by the effective point spread function (PSF). Let us, therefore, prefer to call it the PSF scale parameter $a$. $\theta$ is the angular position of the point where the intensity distribution is measured on the observational plane.

The normalized intensity distribution (maxima at $u_1 = 0, u_2 = 0$) for a binary star system on the detector plane can be modeled as  
\begin{equation}
I(u_1,u_2)=
\frac{\left[2\frac{J_1(u_1)}{u_1}\right]^2
+b\left[2\frac{J_1(u_2)}{u_2}\right]^2}
{1+b},
\label{eqn:int}
\end{equation}
where $b = I_{2}/I_{1}$ denotes the brightness ratio of the two stellar components and
\begin{equation}
    u_2 = \frac{\pi (\theta + \Delta \theta)}{a}
\end{equation}
$\Delta \theta$ stands for the angular separation between the stars and is resolved into Cartesian components according to
\begin{eqnarray}
 \Delta \theta_x &=& \Delta \theta \, \sin(\phi), \\
  \Delta \theta_y &=& \Delta \theta \, \cos(\phi),
\end{eqnarray}
where $\phi$ is the position angle of the notional secondary with respect to the primary of the binary system on the sky in degrees from North through East. In this framework, the key parameters to be estimated include the brightness ratio ($b$), the angular separation ($\Delta \theta$), the position angle ($\phi$), and the PSF scale parameter ($a$). These parameters are inferred by fitting the modeled AS to the observed data using a Bayesian approach, thereby enabling the recovery of binary information even when the secondary peaks in the autocorrelation are fainter than the primary peak.

In the absence of any atmospheric turbulence, the observed intensity distribution of binaries should follow equation~\ref{eqn:int}. However, in the presence of our atmosphere, it becomes distorted and blurred, resulting in a chaotic distribution. This situation can be resolved, and the noise can be canceled, using a pipeline based on FT techniques as discussed in section \ref{sec:FT} and outlined below. 

We obtain the FT ($\mathcal F$) of each speckle frame $I_i$
\begin{equation}
    S_i = \mathcal{F}\{ I_i(u_1, u_2) \}
\end{equation}
Further, the average is calculated over the FT of $N$ frames. This procedure returns the AS of the signal
\begin{equation}
A({s}) = \frac{1}{N} \sum_{i=1}^{N} \left|{S_i}
\right|.
\label{equ:model}
\end{equation}
This equation~\ref{equ:model} is modeled using Bayesian inference to obtain the probability distribution of the parameters, where $s \in \left\{b,\Delta\theta,\phi,a\right\}$ represents collectively the set of key parameters of the stellar system in the sky to be estimated using Bayesian inference.

\subsection{The Prior Information}
Before setting the likelihood, we set the prior range of parameters for each binary. It can incorporate physical constraints such as brightness ratios, realistic ranges for separations, or previously known binary information. We denote it by $P(\omega)$, where $\omega$ denotes the adopted uniform range for a given parameter set $s \in \left\{b,\Delta\theta,\phi,a\right\}$.
To avoid constraining the Bayesian inference using previously known binary parameters, uniform and generalized prior ranges are adopted for both binary systems. 

The adopted prior ranges are based on the physical interpretation of the convention that the primary component is the brighter star, the binary parameters, the expected angular scales of the observed systems, and the observational characteristics of the 3.6-m DOT. In particular, the brightness ratio $b$ is restricted to 0–1 because the primary component is defined as the brighter component; the angular separation $\Delta\theta$ is allowed to vary between 0 and 3 arcsec to encompass the range of interest; and the position angle $\phi$ is restricted to $0^\circ$–$180^\circ$ because of the intrinsic $180^\circ$ ambiguity associated with the binary intensity distribution. The PSF scale parameter is assigned a broad range of 0–2 arcsec, which is specific to the observations obtained with the 3.6-m DOT at the taken wavelength.

\subsection{The Likelihood and Estimation of Parameters}
The next step is to estimate the underlying parameters of the binary system by fitting the model to the observed data through the AS (defined in equation~\ref{equ:model}). We assume that the AS derived from the observed speckle patterns, denoted \(A_{\mathrm{obs}}(s)\), is contaminated by additive Gaussian noise characterized by a standard deviation $\sigma_s$. A global estimate of \(\sigma_s\) is obtained using a standard-deviation-based approach applied to the intensity values of the AS image.

In the Bayesian framework, the likelihood function quantifies the probability of obtaining the measured AS given the model parameters $\omega$. The likelihood function is defined according to
\begin{equation}
    \ln P \left ( \{A_{obs}(s)\}| \omega \right ) = -{\textstyle\frac12 \sum_s
	\frac{( A_{obs}(s) - A(s, \omega))^2}{\sigma_s^{2}}}
    \label{equ:ln_likli}
\end{equation}
We use equation~\ref{equ:ln_likli} for sampling and parameter estimation, which is carried out by maximizing the posterior probability $P(\omega \, | \, A_{obs}(s))$. It combines the likelihood with prior information on the parameters through Bayes’ theorem:  
\begin{eqnarray}
    P(\omega \, | \, A_{obs}(s)) \propto P\left( A_{obs}(s)|\omega\right) \, P(\omega).
\end{eqnarray} 

We use the \texttt{dynesty} package to sample the data and to estimate the parameters \citep{speagle2020dynesty}. This package provides not only the most probable values of the binary parameters but also robust estimates of their uncertainties and correlations. The resulting posterior distributions allow us to quantify confidence in measurements of brightness ratio, angular positions, position angle, and PSF scaling. 
\section{Speckles Observations of Binaries}
We use the 3.6-m DOT to observe the speckle pattern of binary systems. The telescope consists of a Ritchey–Chrétien optical configuration with an effective focal ratio of $f/9$ \citep{2018BSRSL..87...29K}. It provides robust observational capabilities across the optical and near-infrared (NIR) regimes \citep{2018BSRSL..87...29K}. It has three instrument ports: a main (axial) port and two side ports. For the present study, a scientific complementary metal–oxide–semiconductor (sCMOS) camera (ANDOR Marana\footnote{\url{https://andor.oxinst.com/products/scmos-camera-series/marana-scmos}}) was mounted at the main port of the DOT to facilitate SI observations. It enables high-speed image acquisition on millisecond timescales, which is crucial for effectively freezing atmospheric turbulence and, consequently, for recovering diffraction-limited spatial information. The sCMOS camera consists of a detector array of $2048 \times 2048$ pixels with a physical pixel size of 6.5~$\mu$m, yielding an effective field of view of approximately $1.4^\prime \times 1.4^\prime$ on the sky. 
The detector was oriented so that the obtained images have North (N) up and East (E) to the left, as shown in Figures.~\ref{fig:Ori} and ~\ref{fig:Ari}. The readout noise of this sCMOS camera is $\sim$ $1.6 e^-$ at a readout speed of $\sim$ 310 MHz, and the gain is $\sim$ 1 e$^-$/ADU. The details on characterization and setup of the instrument will be presented in \citet[][in preparation]{tsk2025}.

SI observations were carried out for two binaries, 52 Ori and 10 Ari, using short-exposure imaging sequences on 4 October 2024, in median atmospheric seeing conditions of around $1''$. Observations of 52 Ori were carried out without a filter, whereas 10 Ari were observed through the SDSS g-band filter (central wavelength $\sim$ 470 nm and a bandwidth of $\sim$ 128 nm). The Atmospheric Dispersion Correctors were not applied to these observations because the targets were observed near zenith. To account for the dynamic nature of atmospheric turbulence, we employed a multi-exposure acquisition strategy. This approach ensured that the time-varying atmospheric speckles were sufficiently sampled to reconstruct the underlying interference patterns. Different exposure times also help compare results and parameter uncertainties, as well as atmospheric information. 
A total of 10,000 frames of 2 ms (52 Ori), 1,000 frames of 10 ms (52 Ori and 10 Ari), and 1,000 frames of 100 ms (10 Ari) exposure time were taken to optimize the signal-to-noise ratio (SNR).

In a preliminary analysis, the centroid motion and speckle contrast were computed for all frames. Among the datasets, the sequence of 10,000 exposures of 52 Ori, each with a 2 ms exposure time, exhibited systematic drifts in image centroid positions and variations in speckle contrast. However, three intervals (frames 1--1500, 2501--3500, and 8501--9500) showed relatively better speckle contrast and centroid stability, possibly reflecting variations in atmospheric conditions during the observations. Therefore, only the first 1500 frames were retained for the subsequent analysis of the 2 ms exposures of 52 Ori. This selection procedure is similar in principle to Lucky Imaging, in which frames with better image quality are preferentially selected \citep{fried1978probability}, although no formal Lucky Imaging algorithm was implemented. 
The same procedure was performed for the remaining datasets. In these cases, variations in centroid positions and speckle contrast appeared random, with no discernible systematic trends. Consequently, all available frames from these datasets were retained for further analysis.

\section{The Target Sources and Parameter Estimation}
\label{sec:result}
Before we subject the acquired speckle data on 52~Ori and 10~Ari to the procedure outlined in Section \ref{sec:theory}, we reduce the data by subtracting the mean photon count from each of the speckle frames. Next, we compute the FT of these resulting short-exposure frames and obtain their corresponding power spectra. These individual power spectra are averaged across multiple frames, and an inverse FT is computed to obtain the source's autocorrelation. The above steps of Fourier processing of data were conducted using a custom Python-based pipeline built upon standard scientific libraries, including \texttt{NumPy, matplotlib, dynesty, SciPy, corner}, and \texttt{Astropy}.

For the scaling of the autocorrelation image, we convert each pixel of the sCMOS camera using the traditional method $(\rm X = \rm x \, \times \rm pixel \, \rm scale, \rm Y = \rm y \, \times \rm pixel \, \rm scale)$, where $(\rm x, \rm y)$ are the pixel numbers and pixel scale = $0.041''$. We measure the projected distance from the central peak to one secondary peak along the RA and Dec, which allows us to calculate the angular separation $\Delta \theta = \sqrt{\Delta \theta_x^2 + \Delta \theta_y^2}$ and position angle from North $\phi = \tan^{-1}(\frac{\Delta \theta_x}{\Delta \theta_y})$.

\begin{table*}[t]
\centering
\caption{Binary star systems and parameters obtained using the autocorrelation method. The columns, from left to right, are for the binary system, exposure time ($\alpha$), number of observed frames, observation wavelength and bandwidth ($\lambda \pm \Delta\lambda$), angular separation ($\Delta\theta$), position angle from North through East ($\phi$), and with $180^\circ$ ambiguity.}
\label{table:auto}

\setlength{\tabcolsep}{4pt}

\begin{tabular*}{\textwidth}{@{\extracolsep{\fill}}ccccccc@{}}
\toprule
\textbf{Binaries} &
\textbf{$\bm{\alpha}$ (ms)} &
\textbf{No. of Frames} &
\textbf{$\bm{\lambda} \pm \bm{\Delta\lambda}$ (nm)} &
\textbf{$\bm{\Delta\theta''}$} &
\textbf{$\bm{\phi^\circ}$} &
\textbf{$\bm{180^\circ+\phi^\circ}$}\\
\midrule

\multirow{2}{*}{52 Ori}
& 2  & 1500 & 400--900       & 0.94  & 39.7  & 219.7 \\
& 10 & 1000  & 400--900       & 0.94  & 39.7  & 219.7 \\

\multirow{2}{*}{10 Ari}
& 10  & 1000 & $470 \pm 128$ & 1.518 & 165.1 & 345.1\\
& 100 & 1000 & $470 \pm 128$ & 1.518 & 165.1 & 345.1 \\

\bottomrule
\end{tabular*}
\end{table*}

\begin{table*}[t]
\centering
\caption{Binary star systems and estimated parameters obtained using Bayesian inference while modeling the AS of speckles for two different exposure times. The columns are for exposure time ($\alpha$), observation wavelength and bandwidth ($\lambda \pm \Delta\lambda$), standard deviation ($\sigma_s$), brightness ratio ($b$), angular separation ($\Delta\theta$), position angle ($\phi$) with $180^\circ$ ambiguity, and PSF scale parameter ($a$).}
\label{table:binary}

\setlength{\tabcolsep}{4pt}

\begin{tabular*}{\textwidth}{@{\extracolsep{\fill}}cccccccc@{}}
\toprule
\textbf{Binaries} &
\textbf{$\bm{\alpha}$ (ms)} &
\textbf{$\bm{\lambda} \pm \bm{\Delta\lambda}$ (nm)} &
\textbf{$\bm{\sigma_s}$} &
\textbf{$\bm{b}$} &
\textbf{$\bm{\Delta\theta''}$} &
\textbf{$\bm{\phi^\circ}$} &
\textbf{$\bm{a''}$}\\
\midrule

\multirow{2}{*}{52 Ori}
& 2  & 400--900      & 0.019 & $0.63 \pm 0.02$ & $1.069 \pm 0.010$ & $39.5 \pm 0.7$ & $1.190 \pm 0.009$ \\
& 10 & 400--900      & 0.020 & $0.76 \pm 0.02$ & $1.055 \pm 0.009$ & $41.0 \pm 0.6$ & $1.195 \pm 0.007$ \\

\multirow{2}{*}{10 Ari}
& 10  & $470 \pm 128$ & 0.010 & $0.09 \pm 0.01$ & $1.527 \pm 0.008$ & $165.9 \pm 0.3$ & $0.112 \pm 0.001$ \\
& 100 & $470 \pm 128$ & 0.010 & $0.11 \pm 0.008$ & $1.527 \pm 0.006$ & $165.8 \pm 0.3$ & $0.158 \pm 0.001$ \\

\bottomrule
\end{tabular*}
\end{table*}

While the conventional autocorrelation procedure yields a clear visualization of the binary nature of 52 Ori, the autocorrelation image of 10 Ari is further improved by incorporating the AS; however, neither approach provides the corresponding probabilistic distributions of the underlying parameters. To overcome this, we employ Bayesian inference to model the AS and estimate system parameters with improved accuracy (see Section \ref{sec:model}). One of the key inputs of the Bayesian Model, the global standard deviation $\sigma_s$, is obtained using the traditional method on the AS image. The values of $\sigma_s$ are reported in Table~\ref{table:binary} for both binaries at each exposure time.

\subsection{52 Ori}
\subsubsection{Autocorrelation}
52 Ori is a well-known binary star system composed of two main-sequence stars located at a distance of approximately 479.41 light-years from the Sun \citep{2021AJ....161..147B}. Figure~\ref{fig:Ori} illustrates one of the representative speckle patterns from 1500 and 1000 frames obtained with exposure times of 2 ms (left panel) and 10 ms (right panel), respectively.

As described in Section~\ref{sec:theory}, the average of the FT of these individual image frames recovers the underlying diffraction features of the binary system. This effect is clearly visible in Figure~\ref{fig:Ori_FT}, where straight interference fringes confirm the binary nature of 52 Ori. The inclination of these fringes with respect to the horizontal axis indicates the position angle of the binary components relative to the telescope’s projected position. 

The autocorrelation map shown in Figure~\ref{fig:Ori_auto} reveals a characteristic three-lobed structure, with a central peak flanked symmetrically by two side peaks, indicating the binary nature of the system. Figure \ref{fig:Ori_auto} also shows an elliptical point spread function (PSF) surrounding the three autocorrelation peaks, probably due to the result of the uncalibrated autocorrelation function. While the absence of reference star observations during the run precludes traditional calibration and direct photometric estimation, our Bayesian inference approach robustly compensates for these effects. By explicitly modeling the AS, this method enables us to extract accurate system parameters even in the absence of direct calibration.

We obtained two sets of speckle observations with exposure times of 2 ms (10,000 frames {\bf reduced to 1,500 frames as explained earlier}) and 10 ms (1,000 frames). The left panel of Fig.~\ref{fig:Ori_auto} for 2 ms exposure using the first 1500 frames, reveals separations of $\Delta \theta_x \approx -0.601''$ and $\Delta \theta_y \approx -0.724''$, yielding a relative separation of $\Delta \theta \approx 0.94''$ and a position angle of $\phi \approx 39.7^\circ$ (or $219.7^\circ$ due to the $180^\circ$ ambiguity). These results are similar to those from the right panel of Fig. \ref{fig:Ori_auto} (for 10 ms exposure and 1,000 frames), which yields consistent values of $\Delta \theta \approx 0.94''$ and $\phi \approx 39.7^\circ$ (or $219.7^\circ$). Table~\ref{table:auto} summarizes the results obtained from autocorrelation for both datasets.

\subsubsection{Bayesian Inference}
Bayesian inference–based model fitting was performed on the AS data of 52 Ori as shown in Fig.~\ref{fig:Ori_auto}. 
The resulting posterior distributions, shown in the corner plots of Fig.~\ref{fig:Ori_para}, provide the inferred parameters for both exposure datasets. The derived parameters, along with their associated uncertainties, are given in Table \ref{table:binary} and are described below.
\begin{itemize}\setlength{\itemsep}{-0.01cm}
    \item The brightness ratio, $b$, is estimated as $0.63 \pm 0.02$ for the 2~ms exposure and $0.76 \pm 0.02$ for the 10~ms exposure. For the unfiltered broadband observations, these ratios correspond to apparent magnitude differences of $\Delta m = 0.49 \pm 0.03$ and $0.30 \pm 0.03$, respectively. The two estimates differ by $\sim 4.5\sigma$, implying a statistically significant dependence on exposure time. This inconsistency may arise from the lack of photometric calibration, variations in atmospheric seeing between the two exposures, or residual systematics associated with the treatment of seeing and noise in the adopted model.
    
    The derived magnitude differences are broadly comparable to the literature value of $\Delta m = 0.41 \pm 0.01$ reported in the $I$-band \citep{roberts2005adaptive}, while the $V$-band value of $\Delta m = 0.066$ \citep{1997yCat.1239....0E} differs substantially.  However, the comparison is not direct because our speckle observations were obtained without a filter, whereas the literature values correspond to specific photometric bands. The effective spectral response of the unfiltered observations can therefore produce a different flux ratio. We therefore treat the unfiltered brightness ratios primarily as integrated modeling parameters covering the entire visible and near-IR ranges. 
    
    \item The binary separation, $\Delta \theta$, is estimated to be $1.069'' \pm 0.010''$ and $1.055'' \pm 0.009''$ for the 2 ms and 10 ms exposures, respectively. These values are marginally larger than those obtained via the autocorrelation analysis (Table~\ref{table:auto}). Previous speckle interferometric observations reported angular separations of $1.159''$ in 1997 \citep{1997yCat.1239....0E}, $1.08''$ in 2003 \citep{roberts2005adaptive}, $1.03''$ in 2009 \citep{scardia2011speckle}, $1.02''$ and $0.96''$ on two nights in February 2015 \citep{maiwald2016double}, and $0.99''$ in 2023 \citep{alzner2022micrometer}). The values are consistent with our results derived from the autocorrelation and Bayesian methods.
    The small deviation from the past values may arise from changes in the components' positions in the binary orbit across different observational epochs, though systematic modeling and observational uncertainties cannot be ruled out.
    
    \item The inferred position angles, $\phi$, are $39.5^\circ$ (or $219.5^\circ$) $\pm 0.6^\circ$ and $41.0^\circ$ (or $221.0^\circ$) $\pm 0.6^\circ$ for the 2~ms and 10~ms exposures, respectively. Accounting for the inherent $180^\circ$ ambiguity in the autocorrelation function, these values are in good agreement with previous determinations of $216^\circ$ \citep{roberts2005adaptive}, $219^\circ$ \citep{scardia2011speckle}, $221^\circ$ and $221.9^\circ$ \citep{maiwald2016double}, and $221.2^\circ$ \citep{alzner2022micrometer}. The consistency with these independent measurements supports the reliability of the inferred position angle from our Bayesian modeling approach.
    
    \item Values of the PSF scale parameter ($a$) are inferred to be $1.190'' \pm 0.009''$ and $1.195'' \pm 0.007''$ for the 2 ms and 10 ms exposures, respectively. The $a$ parameter characterizes the width of the model point-spread function during the observation and reflects the combined effects of the telescope optics and atmospheric turbulence. Consequently, it provides useful information for assessing the image quality under different observing conditions.
\end{itemize}

\subsection{10 Ari}
\subsubsection{Autocorrelation}
10 Ari is a binary star system with an orbital period of approximately 325 years \citep{heintz1996observations}, located at a distance of about 159 light-years \citep{2021AJ....161..147B} from the Sun. The system consists of a bright F-type main-sequence primary star and a fainter secondary companion with apparent R-band magnitudes of $\sim$ 5.5 mag and 7.5 mag, respectively \citep{tokovinin2016speckle}. We recorded speckle patterns of this binary system at two different exposure times, 10 ms and 100 ms, acquiring 1000 data frames for each. Representative speckle patterns for both exposure times are shown in the left and right panels of Fig.~\ref{fig:Ari}. Both datasets undergo Fourier processing, and the square root of their amplitude spectra is shown in Fig.~\ref{fig:Ari_FT}. The fringes are faint, barely visible around the central peak. The lack of a clear, bright interference pattern could be due to the low SNR measured. However, it provides information about the orientation of sources. We further performed an autocorrelation analysis of the PS to estimate the source's relative angular separation and position angle, as shown in the left and right panels of Fig.~\ref{fig:Ari_auto} for 10- and 100 ms exposures, respectively. However, we could not achieve sufficient sensitivity to clearly visualize the secondary peaks around the central one. Therefore, we applied the inverse Fast Fourier transform (IFFT) on the AS to enhance the visibility of the secondary peak in the system, and the results are shown in the left and right panels of Fig.~\ref{fig:Ari_auto_PS} for 10 ms and 100 ms exposures, respectively. Following the same procedure as performed for  52 Ori, the $\Delta \theta$  and $\phi$ for 10 Ari are estimated to be  $\approx 1.518''$ and $165.1^\circ$ or $345.1^\circ$, respectively, and are mentioned in Table~\ref{table:auto}.

\subsubsection{Bayesian Inference} 
The AS of 10 Ari shown in Fig.~\ref{fig:Ari_FT} is detectable, although the binary fringes become visible only after taking the square root of its pixels. Since the fringes are relatively faint, we model the inverse Fourier transform of the AS (shown in Fig.~\ref{fig:Ari_auto_PS}) using Bayesian inference. Now IFFT of AS represents $A_{obs}(s)$ in Eq.~\ref{equ:ln_likli} and the corresponding model, $A(s)$, is defined as
\begin{equation}
    A(s) = \mathcal{F}^{-1}\left\{
    \frac{1}{N} \sum_{i=1}^{N} \left|S_i\right|
    \right\},
    \label{equ:model_PS}
\end{equation}
Before applying Bayesian inference to $A_{\rm obs}(s)$, we remove the broad, smooth PSF surrounding the central peak using unsharp masking with a Gaussian smoothing kernel. This procedure suppresses the low-spatial-frequency contribution associated with the broad PSF while retaining the localized binary signature. Without this treatment, the extended central seeing contribution dominates the likelihood, limiting posterior exploration of $\Delta\theta$ and $\phi$ beyond the broad PSF. Unsharp masking was therefore applied only to the 10 Ari datasets, where the fainter companion and its secondary peaks were located outside the central broad PSF. For 52 Ori, Bayesian inference was performed directly on the AS without PSF removal. A consistent PSF treatment for both systems will be explored in the development of a standardized SI pipeline.

Fig.~\ref{fig:Ari_para} shows the posterior distributions obtained using the Bayesian inference framework on Equation \ref{equ:model_PS} for both exposures associated with the speckle images in Fig.~\ref{fig:Ari}. The inferred parameters are given in Table \ref{table:binary} and described below.

\begin{itemize}\setlength{\itemsep}{-0.01cm}
\item The inferred values of brightness ratio $b$ for 10 Ari are $0.091 \pm 0.010$ and $0.110 \pm 0.008$ for the 10 ms and 100 ms exposure datasets, corresponding to g-band magnitude differences between the binary components of $2.60 \pm 0.12$ mag and  $2.40 \pm 0.08$ mag, respectively. Both estimates agree well with the reported magnitude differences of 2.3 mag in the R band and 2.18 mag in the I band for this binary system \citep{tokovinin2016speckle}, indicating that observations obtained with the filter yield a more reliable estimate of the brightness ratio and, therefore, the apparent magnitude.

\item The inferred $\Delta \theta$ of $1.527'' \pm 0.008''$ and $1.527'' \pm 0.006''$ for the 10 ms and 100 ms exposure datasets, respectively. The previous speckle interferometric determinations $\Delta \theta$ are  approximately $1.043''$ in 1997 \citep{1997yCat.1239....0E}, $1.216''$ in 2004 \citep{10.1111/j.1365-2966.2006.10035.x}, $1.419''$ in 2016 \citep{wasson2018speckle}, and $1.410''$ in 2017 \citep{harshaw2018measurements}. Differences among measurements obtained at different epochs may partly arise from the orbital motion of the binary system.

\item The $\phi$, is estimated to be $165.9^\circ$ or $345.9^\circ \pm 0.3^\circ$ for the 10 ms exposure dataset and $165.8^\circ$ or $345.8^\circ \pm 0.3^\circ$ for the 100 ms exposure dataset. The values inferred from both datasets are consistent with the position angle of approximately $345.1^\circ$ obtained via inverse Fourier transformation of the AS. Previous determinations of $\phi$ include $\sim 328^\circ$ in 1997 \citep{1997yCat.1239....0E}, $\sim 343^\circ$ in 2017 \citep{harshaw2018measurements}, and $344^\circ$ in 2016 \citep{wasson2018speckle}.

\item The PSF scale parameter, $a$, is estimated to be $0.112'' \pm 0.001''$ and $0.158'' \pm 0.001''$ for the 10 ms and 100 ms exposure datasets, respectively, on the g-band filter. Although the PSF scale parameter provides useful information about the atmospheric seeing conditions, a detailed analysis of atmospheric turbulence is beyond the scope of this work.
\end{itemize}
\section{Discussion and Conclusions}
In this work, we present a successful implementation of SI on the 3.6-m DOT, demonstrating the telescope's capability to resolve binary systems with bright companions using FT and autocorrelation techniques. We analyzed white-band observations of 52 Ori, consisting of 1,500 frames with a 2 ms exposure per frame and 1,000 frames with a 10 ms exposure per frame. Additionally, we analyzed $g$-band observations of 10 Ari, comprising 1,000 frames each at 10 ms and 100 ms per frame.
The autocorrelation method provided reliable measurements of the relative separation and position angle for both systems. Although deriving robust probability distributions directly from the autocorrelation method remains challenging in this study, applying Bayesian inference to model the AS and its Fourier transform of speckle patterns shows significant promise for estimating binary parameters with high precision. This approach yielded posterior distributions for the brightness ratio, angular separation, position angle, and PSF scale parameter.

Although the AS provides a useful means of identifying and modeling binary fringe patterns, the nonlinear modulus operation modifies the statistical properties of Fourier-domain noise. Consequently, the use of a Gaussian likelihood and its influence on the inferred parameter uncertainties require further investigation. 
Furthermore, for 10 Ari, we explored the inverse Fourier transform of the AS to enhance the visibility of the faint companion. This representation is not equivalent to the conventional autocorrelation and may influence the recovered brightness ratio and associated uncertainties. The different pre-processing procedures adopted for the two binary systems, including contrast-based frame selection for the 2 ms observations of 52 Ori and additional unsharp masking for 10 Ari, also reflect the exploratory nature of the present implementation.

Future development will focus on establishing a uniform processing framework for all observational datasets. In particular, systematic comparisons of PS and AS modeling, statistically appropriate noise likelihoods, objective frame-selection criteria, and consistent PSF treatment will be essential. These improvements will be important for establishing a reliable and reproducible SI pipeline.

For 52 Ori, mean parameters were derived from 2 ms and 10 ms exposures, yielding a magnitude difference of $\Delta m = 0.39 \pm 0.02$ mag, an angular separation of $\Delta \theta \approx 1.062'' \pm 0.007''$, and a position angle of $\phi \approx 220.3^\circ \pm 0.4^\circ$. For 10 Ari, analysis of 10 ms and 100 ms exposures produced mean values of $\Delta m = 2.55 \pm 0.07$ mag, $\Delta \theta = 1.527'' \pm 0.005''$, and $\phi = 345.9^\circ \pm 0.2^\circ$. The corresponding PSF scaling estimates were $1.192'' \pm 0.006''$ for 52 Ori in the white band and $0.135'' \pm 0.001''$ for 10 Ari in the $g$ band, respectively, and can provide further information about the atmospheric seeing at the DOT site. These results are consistent with previously reported measurements, validating both the autocorrelation analysis and the Bayesian modeling approach. Even though the paper does not demonstrate diffraction-limited imaging using SI, both analyzed binaries have separations well above the telescope’s diffraction limit and are approximately comparable to the median seeing. It establishes a robust foundation for high-spatial-resolution observations of single, binary, and multiple stellar systems using telescopes like DOT.

Our Bayesian analysis further demonstrates that the exposure time and number of frames can influence the precision of the recovered binary parameters. Increasing the integration time can enhance the per-frame SNR, thereby potentially improving constraints on inferred quantities such as angular separation and magnitude difference. Nevertheless, the influence of exposure duration is not governed exclusively by the higher signal level. When the integration time approaches or exceeds the atmospheric coherence time under the relevant observing conditions, temporal averaging of the atmospheric speckle pattern can occur. This averaging may partially blur the speckle morphology, yielding an intensity distribution that increasingly approximates that of standard long-exposure imaging. Accordingly, the uncertainties of the recovered parameters are expected to depend jointly on the per-frame SNR, the total number of acquired frames, and the degree of temporal averaging of the atmospheric speckle pattern. Conversely, increasing the number of short-exposure frames can compensate for the lower signal per frame by improving the statistical signal-to-noise ratio of the ensemble-averaged AS. These results highlight the importance of selecting an appropriate combination of exposure time and frame statistics for reliable parameter recovery in speckle-interferometric observations.

The present analysis also indicates that using a narrower optical filter improves both photometric and astrometric precision. Specifically, the magnitude difference measured for 10 Ari using the SDSS $g$-band exhibits higher precision than that obtained for 52 Ori, which was observed in white light. This improvement is likely attributable to the reduced spectral bandwidth of the SDSS $g$-band, which minimizes chromatic smearing and produces more coherent and less elongated speckles, enabling more accurate photometric and astrometric measurements.

Overall, this work demonstrates the feasibility of implementing SI as a backend instrument on the 3.6-m DOT. With optimized observing strategies, including appropriate exposure times, sufficient frame statistics, and suitable filters, SI can be a highly effective technique for achieving diffraction-limited resolution.

\section*{Acknowledgments} 
We sincerely thank the anonymous referee for the constructive comments and valuable suggestions, which have significantly improved the quality of this manuscript. The authors thank the Surya Cluster facilities provided by ARIES for computing performance. One of the authors (SS) gratefully acknowledges the local hospitality provided by the Inter-University Center for Astronomy and Astrophysics (IUCAA), Pune, India, under its Visiting Associate Program during the preparation of this manuscript.

\bibliography{main}

\begin{thebibliography}{}
\expandafter\ifx\csname natexlab\endcsname\relax\def\natexlab#1{#1}\fi
\providecommand{\url}[1]{\href{#1}{#1}}
\providecommand{\dodoi}[1]{doi:~\href{http://doi.org/#1}{\nolinkurl{#1}}}
\providecommand{\doeprint}[1]{\href{http://ascl.net/#1}{\nolinkurl{http://ascl.net/#1}}}
\providecommand{\doarXiv}[1]{\href{https://arxiv.org/abs/#1}{\nolinkurl{https://arxiv.org/abs/#1}}}

% type= article
\bibitem[{A. Alzner(2023)Alzner}]{alzner2022micrometer}
Alzner, A. 2023, \bibinfo{title}{MICROMETER MEASURES OF DOUBLE STARS WITH A
  32.5-CM F19 CASSEGRAIN AND A MECA-PRECIS DOUBLE IMAGE MICROMETER FROM 2021.14
  TO 2023.14,} MICROMETRIC MEASURES OF DOUBLE STARS IN 2022, 6.
\newblock \url{http://www.sterngucker.eu/dssc31.pdf#page=8}

% type= article
\bibitem[{C.~A.~L. {Bailer-Jones} {et~al.}(2021){Bailer-Jones}, {Rybizki},
  {Fouesneau}, {Demleitner}, \& {Andrae}}]{2021AJ....161..147B}
{Bailer-Jones}, C.~A.~L., {Rybizki}, J., {Fouesneau}, M., {Demleitner}, M., \&
  {Andrae}, R. 2021, \bibinfo{title}{{Estimating Distances from Parallaxes. V.
  Geometric and Photogeometric Distances to 1.47 Billion Stars in Gaia Early
  Data Release 3},} \aj, 161, 147, \dodoi{10.3847/1538-3881/abd806}

% type= article
\bibitem[{J. Butters \& J. Leendertz(1971)Butters \&
  Leendertz}]{butters1971double}
Butters, J., \& Leendertz, J. 1971, \bibinfo{title}{A double exposure technique
  for speckle pattern interferometry,} Journal of Physics E: Scientific
  Instruments, 4, 277, \dodoi{10.1088/0022-3735/4/4/004}

% type= inbook
\bibitem[{J.~C. Dainty(1975)Dainty}]{Dainty1975}
Dainty, J.~C. 1975, Stellar Speckle Interferometry, ed. J.~C. Dainty (Berlin,
  Heidelberg: Springer Berlin Heidelberg), 255--280,
  \dodoi{10.1007/978-3-662-43205-1_7}

% type=
\bibitem[{ {Esa}(1997){Esa}}]{1997yCat.1239....0E}
{Esa}. 1997, {VizieR Online Data Catalog: The Hipparcos and Tycho Catalogues
  (ESA 1997)}, \url{https://ui.adsabs.harvard.edu/abs/1997yCat.1239....0E}

% type= article
\bibitem[{D.~L. Fried(1978)Fried}]{fried1978probability}
Fried, D.~L. 1978, \bibinfo{title}{Probability of getting a lucky
  short-exposure image through turbulence,} Journal of the Optical Society of
  America, 68, 1651, \dodoi{10.1364/JOSA.68.001651}

% type= article
\bibitem[{R.~J. Glauber(1963)Glauber}]{Glauber1963PhotonCorrelations}
Glauber, R.~J. 1963, \bibinfo{title}{Photon Correlations,} Physical Review
  Letters, 10, 84, \dodoi{10.1103/PhysRevLett.10.84}

% type= article
\bibitem[{L.~I. Goldfischer(1965)Goldfischer}]{goldfischer1965autocorrelation}
Goldfischer, L.~I. 1965, \bibinfo{title}{Autocorrelation function and power
  spectral density of laser-produced speckle patterns,} Journal of the optical
  society of america, 55, 247, \dodoi{10.1364/JOSA.55.000247}

% type= book
\bibitem[{J.~W. Goodman(2007)Goodman}]{goodman2007speckle}
Goodman, J.~W. 2007, Speckle phenomena in optics: theory and applications
  (Roberts and Company Publishers), \dodoi{10.1117/3.2548484}

% type= article
\bibitem[{R. {Hanbury Brown} \& R.~Q. {Twiss}(1957){Hanbury Brown} \&
  {Twiss}}]{1957RSPSA.242..300B}
{Hanbury Brown}, R., \& {Twiss}, R.~Q. 1957, \bibinfo{title}{{Interferometry of
  the Intensity Fluctuations in Light. I. Basic Theory: The Correlation between
  Photons in Coherent Beams of Radiation},} Proceedings of the Royal Society of
  London Series A, 242, 300, \dodoi{10.1098/rspa.1957.0177}

% type= book
\bibitem[{J.~W. {Hardy}(1998){Hardy}}]{hardy1998adaptive}
{Hardy}, J.~W. 1998, {Adaptive Optics for Astronomical Telescopes}, Vol.~16
  (Oxford university press).
\newblock \url{https://ui.adsabs.harvard.edu/abs/1998aoat.book.....H}

% type= article
\bibitem[{R. Harshaw(2018)Harshaw}]{harshaw2018measurements}
Harshaw, R. 2018, \bibinfo{title}{Measurements of 427 Double Stars With Speckle
  Interferometry: The Winter/Spring 2017 Observing Program at Brilliant Sky
  Observatory, Part,} Measurements, 14.
\newblock \url{http://www.jdso.org/volume14/number2/Harshaw_284_330.pdf}

% type= article
\bibitem[{W.~I. Hartkopf {et~al.}(1999)Hartkopf, Mason, Gies, ten Brummelaar,
  McAlister, Moffat, Shara, \& Wallace}]{hartkopf1999iccd}
Hartkopf, W.~I., Mason, B.~D., Gies, D.~R., {et~al.} 1999, \bibinfo{title}{ICCD
  Speckle Observations of Binary Stars. XXII. A Survey ofWolf-Rayet
  StarsforClose Visual Companions,} ApJ, 118, 509, \dodoi{10.1086/300943}

% type= article
\bibitem[{W. Heintz(1996)Heintz}]{heintz1996observations}
Heintz, W. 1996, \bibinfo{title}{Observations of Double Stars and New Pairs.
  XVII.,} Astrophysical Journal Supplement v. 105, p. 475, 105, 475,
  \dodoi{10.1086/192324}

% type= article
\bibitem[{E.~P. Horch {et~al.}(2017)Horch, Casetti-Dinescu, Camarata, Bidarian,
  Altena, Sherry, Everett, Howell, Ciardi, Henry, Nusdeo, \&
  Winters}]{horch2017}
Horch, E.~P., Casetti-Dinescu, D.~I., Camarata, M.~A., {et~al.} 2017,
  \bibinfo{title}{Observations of binary stars with the differential speckle
  survey instrument. VII. Measures from 2010 September to 2012 February at the
  WIYN telescope,} ApJ, 153, 212, \dodoi{10.3847/1538-3881/aa6749}

% type= article
\bibitem[{S.~B. Howell {et~al.}(2025)Howell, Ciardi, Clark, Hope, Littlefield,
  \& Furlan}]{Howell_2025}
Howell, S.~B., Ciardi, D.~R., Clark, C.~A., {et~al.} 2025, \bibinfo{title}{The
  Probable Direct-imaging Detection of the Stellar Companion to Betelgeuse,}
  The Astrophysical Journal Letters, 988, L47, \dodoi{10.3847/2041-8213/adeaaf}

% type= article
\bibitem[{S.~B. Howell {et~al.}(2024)Howell, Martinez, Hope, Ciardi, Jefferies,
  Baron, \& Lund}]{howell2024high}
Howell, S.~B., Martinez, A.~O., Hope, D.~A., {et~al.} 2024,
  \bibinfo{title}{High-contrast, high-angular-resolution optical speckle
  imaging: uncovering hidden stellar companions,} The Astronomical Journal,
  167, 258, \dodoi{10.3847/1538-3881/ad3df2}

% type= article
\bibitem[{A. Khintchine(1934)Khintchine}]{khintchine1934korrelationstheorie}
Khintchine, A. 1934, \bibinfo{title}{Korrelationstheorie der station{\"a}ren
  stochastischen Prozesse,} Mathematische Annalen, 109, 604,
  \dodoi{10.1007/BF01449156}

% type= article
\bibitem[{B. Kumar {et~al.}(2018)Kumar, Omar, Maheswar, Pandey, Sagar, Uddin,
  Sanwal, Bangia, Kumar, Yadav, {et~al.}}]{2018BSRSL..87...29K}
Kumar, B., Omar, A., Maheswar, G., {et~al.} 2018, \bibinfo{title}{{3.6-m
  Devasthal Optical Telescope Project: Completion and first results},} Bulletin
  de la Societe Royale des Sciences de Liege, 87, 29.
\newblock \url{http://prints.iiap.res.in/handle/2248/6829}

% type= article
\bibitem[{T.~S. Kumar {et~al.}(2026)Kumar, Kumar, Ray, \& Shah}]{tsk2025}
Kumar, T.~S., Kumar, B., Ray, S., \& Shah, M. 2026,
  \bibinfo{title}{Implementation of Speckle Interferometry Using a High-Speed
  CMOS Detector on the 3.6 m DOT,} (Manuscript in preparation)

% type= article
\bibitem[{A. Labeyrie(1970)Labeyrie}]{labeyrie1970attainment}
Labeyrie, A. 1970, \bibinfo{title}{Attainment of diffraction limited resolution
  in large telescopes by Fourier analysing speckle patterns in star images,}
  Astronomy and Astrophysics, Vol. 6, p. 85 (1970), 6, 85.
\newblock \url{https://ui.adsabs.harvard.edu/abs/1970A&A.....6...85L}

% type= article
\bibitem[{A.~W. Lohmann {et~al.}(1983)Lohmann, Weigelt, \&
  Wirnitzer}]{lohmann1983speckle}
Lohmann, A.~W., Weigelt, G., \& Wirnitzer, B. 1983, \bibinfo{title}{Speckle
  masking in astronomy: triple correlation theory and applications,} Applied
  Optics, 22, 4028, \dodoi{10.1364/AO.22.004028}

% type= article
\bibitem[{M.~O. Maiwald(2016)Maiwald}]{maiwald2016double}
Maiwald, M.~O. 2016, \bibinfo{title}{Double Star Observations with a 150mm
  Refractor in 2015,} Journal of Double Star Observations, 12, 437.
\newblock \url{https://ui.adsabs.harvard.edu/abs/2016JDSO...12..437M}

% type= article
\bibitem[{H.~A. McAlister(1977)McAlister}]{mcalister1977speckle}
McAlister, H.~A. 1977, \bibinfo{title}{Speckle interferometric measurements of
  binary stars. I,} ApJ, 215, 159, \dodoi{10.1086/155343}

% type= article
\bibitem[{H.~A. McAlister {et~al.}(1989)McAlister, Hartkopf, Sowell,
  Dombrowski, \& Franz}]{1989AJ.....97..510M}
McAlister, H.~A., Hartkopf, W.~I., Sowell, J.~R., Dombrowski, E.~G., \& Franz,
  O.~G. 1989, \bibinfo{title}{{ICCD Speckle Observations of Binary Stars. IV.
  Measurements During 1986-1988 from the Kitt Peak 4-m Telescope},} \aj, 97,
  510, \dodoi{10.1086/115001}

% type= article
\bibitem[{L.~C. Roberts {et~al.}(2005)Roberts, Turner, Bradford, Brummelaar,
  Oppenheimer, Kuhn, Whitman, Perrin, \& Graham}]{roberts2005adaptive}
Roberts, Jr, L.~C., Turner, N.~H., Bradford, L.~W., {et~al.} 2005,
  \bibinfo{title}{Adaptive optics photometry and astrometry of binary stars,}
  The Astronomical Journal, 130, 2262, \dodoi{10.1086/491586}

% type= article
\bibitem[{R. Sagar {et~al.}(2019)Sagar, Kumar, \& Omar}]{sagar20193}
Sagar, R., Kumar, B., \& Omar, A. 2019, \bibinfo{title}{The 3.6 metre Devasthal
  Optical Telescope,} Current Science, 117, 365.
\newblock \url{https://www.jstor.org/stable/27138269}

% type= article
\bibitem[{R. Sagar {et~al.}(2020)Sagar, Kumar, \& Sharma}]{Sagar2020}
Sagar, R., Kumar, B., \& Sharma, S. 2020, \bibinfo{title}{Observations with the
  3.6-meter Devasthal optical telescope,} Journal of Astrophysics and
  Astronomy, 41, 33, \dodoi{10.1007/s12036-020-09652-9}

% type= article
\bibitem[{S. Saha {et~al.}(1999)Saha, Sudheendra, Chandra, \&
  Chinnappan}]{saha1999development}
Saha, S., Sudheendra, G., Chandra, A.~U., \& Chinnappan, V. 1999,
  \bibinfo{title}{Development of a Speckle Interferometer and the Measurement
  of Fried's Parameter (ro) at the Telescope Site,} Experimental Astronomy, 9,
  39, \dodoi{10.1023/A:1008057218328}

% type= article
\bibitem[{S.~K. {Saha} {et~al.}(2002){Saha}, {Chinnappan}, {Yeswanth}, \&
  {Anbazhagan}}]{saha2002speckle}
{Saha}, S.~K., {Chinnappan}, V., {Yeswanth}, L., \& {Anbazhagan}, P. 2002,
  \bibinfo{title}{Speckle interferometric observations of close binary stars.,}
  Bulletin of the Astronomical Society of India, 30, 677,
  \dodoi{10.48550/arXiv.astro-ph/0209424}

% type= article
\bibitem[{M. Scardia {et~al.}(2011)Scardia, Prieur, Pansecchi, Argyle, \&
  Sala}]{scardia2011speckle}
Scardia, M., Prieur, J.-L., Pansecchi, L., Argyle, R., \& Sala, M. 2011,
  \bibinfo{title}{Speckle observations with PISCO in Merate: X. Astrometric
  measurements of visual binaries in 2009,} Astronomische Nachrichten, 332,
  508, \dodoi{10.1002/asna.201011553}

% type= article
\bibitem[{M. Scardia {et~al.}(2006)Scardia, Prieur, Pansecchi, Argyle, Sala,
  Ghigo, Koechlin, \& Aristidi}]{10.1111/j.1365-2966.2006.10035.x}
Scardia, M., Prieur, J.-L., Pansecchi, L., {et~al.} 2006,
  \bibinfo{title}{Speckle observations with PISCO in Merate – II. Astrometric
  measurements of visual binaries in 2004,} Monthly Notices of the Royal
  Astronomical Society, 367, 1170, \dodoi{10.1111/j.1365-2966.2006.10035.x}

% type= article
\bibitem[{N.~J. Scott {et~al.}(2018)Scott, Howell, Horch, \&
  Everett}]{scott2018nn}
Scott, N.~J., Howell, S.~B., Horch, E.~P., \& Everett, M.~E. 2018,
  \bibinfo{title}{The NN-explore exoplanet stellar speckle imager: Instrument
  description and preliminary results,} Publications of the Astronomical
  Society of the Pacific, 130, 054502, \dodoi{10.1088/1538-3873/aab484}

% type= article
\bibitem[{N.~J. Scott {et~al.}(2021)Scott, Howell, Gnilka, Stephens, Salinas,
  Matson, Furlan, Horch, Everett, Ciardi, {et~al.}}]{scott2021twin}
Scott, N.~J., Howell, S.~B., Gnilka, C.~L., {et~al.} 2021, \bibinfo{title}{Twin
  High-resolution, High-speed Imagers for the Gemini Telescopes: Instrument
  description and science verification results,} Frontiers in Astronomy and
  Space Sciences, 8, 716560, \dodoi{10.3389/fspas.2021.716560}

% type= article
\bibitem[{J.~S. Speagle(2020)Speagle}]{speagle2020dynesty}
Speagle, J.~S. 2020, \bibinfo{title}{dynesty: a dynamic nested sampling package
  for estimating Bayesian posteriors and evidences,} Monthly Notices of the
  Royal Astronomical Society, 493, 3132, \dodoi{10.1093/mnras/staa278}

% type= article
\bibitem[{A. {Tokovinin} \& R. {Cantarutti}(2008){Tokovinin} \&
  {Cantarutti}}]{2008PASP..120..170T}
{Tokovinin}, A., \& {Cantarutti}, R. 2008, \bibinfo{title}{{First Speckle
  Interferometry at SOAR Telescope with Electron-Multiplication CCD},} \pasp,
  120, 170, \dodoi{10.1086/528809}

% type= article
\bibitem[{A. Tokovinin \& E.~P. Horch(2016)Tokovinin \&
  Horch}]{tokovinin2016speckle}
Tokovinin, A., \& Horch, E.~P. 2016, \bibinfo{title}{Speckle interferometry of
  secondary components in nearby visual binaries,} The Astronomical Journal,
  152, 116, \dodoi{10.3847/0004-6256/152/5/116}

% type= article
\bibitem[{A. Tokovinin {et~al.}(2024)Tokovinin, Mason, Mendez, \&
  Costa}]{tokovinin2024speckle}
Tokovinin, A., Mason, B.~D., Mendez, R.~A., \& Costa, E. 2024,
  \bibinfo{title}{Speckle Interferometry at SOAR Telescope in 2023,} The
  Astronomical Journal, 168, 28, \dodoi{10.3847/1538-3881/ad4d56}

% type= book
\bibitem[{R.~K. Tyson \& B.~W. Frazier(2022)Tyson \&
  Frazier}]{tyson2022principles}
Tyson, R.~K., \& Frazier, B.~W. 2022, Principles of adaptive optics (CRC
  press), \dodoi{10.1201/9781003140191}

% type= article
\bibitem[{R. Wasson(2018)Wasson}]{wasson2018speckle}
Wasson, R. 2018, \bibinfo{title}{Speckle Interferometry with the OCA Kuhn 22"
  Telescope,} Journal of Double Star Observations, 14, 223.
\newblock \url{https://ui.adsabs.harvard.edu/abs/2018JDSO...14..223W/abstract}

% type= article
\bibitem[{G. Weigelt \& J. Ebersberger(1986)Weigelt \&
  Ebersberger}]{weigelt1986eta}
Weigelt, G., \& Ebersberger, J. 1986, \bibinfo{title}{Eta Carinae resolved by
  speckle interferometry,} A \& A, 163, L5.
\newblock \url{https://adsabs.harvard.edu/full/1986A%26A...163L...5W}

% type= article
\bibitem[{G. Weigelt \& B. Wirnitzer(1983)Weigelt \&
  Wirnitzer}]{weigelt1983image}
Weigelt, G., \& Wirnitzer, B. 1983, \bibinfo{title}{Image reconstruction by the
  speckle-masking method,} Optics Letters, 8, 389, \dodoi{10.1364/OL.8.000389}

% type= article
\bibitem[{G.~P. {Weigelt}(1977){Weigelt}}]{1977OptCo..21...55W}
{Weigelt}, G.~P. 1977, \bibinfo{title}{{Modified astronomical speckle
  interferometry ``speckle masking''},} Optics Communications, 21, 55,
  \dodoi{10.1016/0030-4018(77)90077-3}

\end{thebibliography}
\bibliographystyle{aasjournalv7}
\end{document}